\documentclass[11pt]{article}

\usepackage[final]{acl}

\usepackage{times}
\usepackage{latexsym}
\usepackage[T1]{fontenc}
\usepackage[utf8]{inputenc}
\usepackage{microtype}
\usepackage{inconsolata}
\usepackage{graphicx}
\usepackage{booktabs}
\usepackage{amsmath}
\usepackage{placeins}
\usepackage{pgfplots}
\usepgfplotslibrary{groupplots}
\pgfplotsset{compat=1.18}

\title{MAGIC-TTS: Fine-Grained Controllable Speech Synthesis with Explicit Local Duration and Pause Control}

\author{
  Jialong Mai \quad Xiaofen Xing\thanks{Corresponding author.} \quad Xiangmin Xu \\
  South China University of Technology
}

\begin{document}
\maketitle

\begin{abstract}
Fine-grained local timing control is still absent from modern text-to-speech systems: existing approaches typically provide only utterance-level duration or global speaking-rate control, while precise token-level timing manipulation remains unavailable. To the best of our knowledge, MAGIC-TTS is the first TTS model with explicit local timing control over token-level content duration and pause. MAGIC-TTS is enabled by explicit token-level duration conditioning, carefully prepared high-confidence duration supervision, and training mechanisms that correct zero-value bias and make the model robust to missing local controls. On our timing-control benchmark, MAGIC-TTS substantially improves token-level duration and pause following over spontaneous synthesis. Even when no timing control is provided, MAGIC-TTS maintains natural high-quality synthesis. We further evaluate practical local editing with a scenario-based benchmark covering navigation guidance, guided reading, and accessibility-oriented code reading. In this setting, MAGIC-TTS realizes a reproducible uniform-timing baseline and then moves the edited regions toward the requested local targets with low mean bias. These results show that explicit fine-grained controllability can be implemented effectively in a high-quality TTS system and can support realistic local timing-editing applications.
\end{abstract}

\section{Introduction}
Reliable fine-grained local timing control is still absent from modern text-to-speech systems. Existing approaches typically provide only utterance-level duration control or global speaking-rate adjustment \citep{zhou2025indextts2,guo2023prompttts,lyth2024natural}, while precise token-level timing manipulation remains difficult to achieve reliably. This limitation makes it difficult to support speech generation scenarios that require controlled pacing, explicit boundary placement, or localized emphasis.

Methods that claim finer duration control are also not yet reliable for practical local manipulation: many are built on autoregressive generation, where free-running rollout makes local timing decisions hard to stabilize \citep{wang2023neural,wang2025maskgct}, or they predict less interpretable intermediate representations rather than explicit local timing targets \citep{ren2021fastspeech2,kim2021vits}. These limitations motivate our choice of a flow-based TTS backbone, where local timing conditions can be injected explicitly into a parallel acoustic generator rather than being entangled with autoregressive rollout \citep{le2023voicebox,ju2024naturalspeech3,chen2024f5tts}.

We present MAGIC-TTS, a TTS model with explicit local timing control over token-level content duration and pause. Our goal is not only to provide local timing controllability, but to make such control highly reliable in practice. To this end, MAGIC-TTS combines explicit token-level duration conditioning with carefully prepared high-confidence duration supervision and training mechanisms that correct zero-value bias and make the model robust to missing local controls. Built on a modern flow-based zero-shot TTS backbone \citep{chen2024f5tts}, this design enables the model to follow local timing instructions while preserving strong uncontrolled synthesis behavior.

Empirically, MAGIC-TTS provides reliable pause control and effective fine-grained duration manipulation. At the same time, in the uncontrolled setting, it preserves natural high-quality synthesis. These results show that fine-grained controllability can be integrated with strong default synthesis behavior in a single TTS system.

\paragraph{Contributions.} Our main contributions are as follows:
\begin{itemize}
    \item To the best of our knowledge, we present MAGIC-TTS, the first TTS model with explicit local timing control over token-level content duration and pause.
    \item We propose a training method for highly reliable local control, combining high-confidence local timing supervision, zero-value correction, and robustness to missing local controls.
    \item We develop a carefully prepared duration-data pipeline for local timing supervision, including cross-validated high-confidence duration datasets that support stable local-control training.
    \item We show that MAGIC-TTS delivers reliable fine-grained local control and supports practical local timing editing in realistic scenarios.
\end{itemize}

\section{Related Work}
\subsection{Zero-Shot Speech Synthesis}

Large-scale zero-shot text-to-speech has rapidly improved the naturalness, robustness, and speaker similarity of synthesized speech. Autoregressive codec language models such as VALL-E formulate TTS as neural codec token prediction conditioned on text and a short acoustic prompt \citep{wang2023neural}. Subsequent systems improve zero-shot generation with stronger generative formulations, including flow or diffusion based speech generation in Voicebox, NaturalSpeech 3, and F5-TTS \citep{le2023voicebox,ju2024naturalspeech3,chen2024f5tts}, as well as LLM-based or codec-token based systems such as CosyVoice and CosyVoice 2 \citep{du2024cosyvoice,du2024cosyvoice2}. These models show that large-scale speech generation can achieve strong default synthesis quality and convincing zero-shot voice cloning.

However, timing is usually learned implicitly in these systems rather than exposed as an explicit local control signal. MaskGCT makes this limitation especially clear, noting that autoregressive systems implicitly model duration but lack duration controllability, while many non-autoregressive systems rely on explicit alignments or phoneme-level duration prediction \citep{wang2025maskgct}. MAGIC-TTS follows the recent success of flow-based zero-shot TTS, but augments it with explicit token-level content-duration and pause conditions, so that local timing can be controlled directly while preserving strong zero-shot synthesis quality.

\subsection{Duration and Prosody Control}

Duration and prosody modeling have long been central to neural TTS. Non-autoregressive systems often predict duration as an internal alignment variable: FastSpeech 2 introduces variance predictors for duration, pitch, and energy, while FastPitch conditions synthesis on predicted pitch contours that can be adjusted for expressive control \citep{ren2021fastspeech2,lancucki2021fastpitch}. VITS further integrates stochastic duration prediction into an end-to-end generative TTS framework \citep{kim2021vits}. These approaches improve synthesis efficiency, stability, and prosodic naturalness, but their duration components are usually optimized as latent or intermediate modeling tools rather than as dependable local controls for users. As discussed in our appendix analysis of Grad-TTS and YourTTS, even directly manipulating such intermediate timing variables does not automatically produce reliable local timing control; in practice, it can also degrade synthesis quality, including both timbre fidelity and intelligibility.

Recent large-scale systems also provide broader duration-related controls. IndexTTS2 targets duration-controlled autoregressive zero-shot TTS by allowing users to specify the number of generated tokens, giving precise control over total speech length \citep{zhou2025indextts2}. MiniMax-Speech demonstrates strong zero-shot voice cloning and extensible control capabilities \citep{zhang2025minimaxspeech}. Style-prompted systems can also influence speaking speed through descriptions such as slow or fast speech \citep{guo2023prompttts,lyth2024natural}. Nevertheless, these controls are primarily utterance-level or style-level: they can change overall duration or speaking rate, but do not specify the duration of a particular token or the local pause at a particular boundary. MAGIC-TTS instead models token-level content duration and pause as explicit numerical conditions, enabling fine-grained local timing manipulation.

\subsection{Controllable Text-to-Speech}

Controllable TTS has expanded from acoustic-factor control toward natural-language and instruction-based control. PromptTTS uses text descriptions to control attributes such as speaker traits, pitch, emotion, volume, and speaking speed \citep{guo2023prompttts}. InstructTTS and natural-language-guided TTS further explore style control with natural language prompts in expressive speech synthesis \citep{yang2024instructtts,lyth2024natural}. SpeechCraft contributes a large-scale expressive speech dataset with natural-language descriptions, reflecting a broader trend toward promptable and semantically rich speech generation \citep{jin2024speechcraft}. Recent surveys also characterize controllable speech synthesis as a fast-growing area driven by diffusion models, LLMs, and natural-language control \citep{xie2025survey}.

These works make controllable speech generation more expressive and easier to specify, but their controls are generally high-level, descriptive, and sentence-level. Such controls are valuable for style, emotion, and global speaking manner, yet they are difficult to evaluate as exact local timing instructions. MAGIC-TTS addresses a complementary form of controllability: explicit numerical control over token-local timing, including both the spoken duration and pause associated with each token. This capability helps fill the gap for speech generation scenarios that require precise local pacing, explicit boundary placement, or localized emphasis.

\section{Method}
\label{sec:method}
\subsection{Method Overview}

MAGIC-TTS is designed to make local timing directly controllable while retaining the natural zero-shot synthesis behavior of a flow-based TTS model. Given a text sequence $\mathbf{y}=(y_1,\ldots,y_N)$ and an acoustic prompt, the model generates a mel-spectrogram continuation through a conditional flow-matching acoustic generator. In addition to the usual text and acoustic conditions, MAGIC-TTS optionally receives a token-aligned timing track
\begin{equation}
    \mathbf{r}_i = (d_i, p_i), \quad i=1,\ldots,N,
\end{equation}
where $d_i$ denotes the content duration of token $y_i$ and $p_i$ denotes the pause associated with this token. Both values are represented in acoustic frames. We treat them as explicit numerical conditions, rather than as latent prosodic variables that must be inferred implicitly from text.

This formulation separates two aspects that are usually entangled in TTS. The acoustic generator remains responsible for producing natural speech from text and prompt speech, while the timing track specifies how local time should be allocated. This separation is important because content duration and pause have different control characteristics. Pause mainly controls boundary regions between spoken units, whereas content duration controls the acoustic realization inside a token. The latter is more sensitive to boundary errors and is easier to weaken if the model learns a stronger shortcut through pause conditions. MAGIC-TTS therefore combines an explicit timing-conditioning module with supervision and training designs that make content duration and pause both usable in practice.

The method has three main components. First, we inject the timing track into the text-side representation of a pretrained flow-based TTS backbone, so local timing can affect generation without changing the flow-matching objective. Second, we construct timing supervision at two levels: large-scale Stable-ts-based labels \citep{jianfch2025stablets} for broad duration-conditioned continued training, and a cross-validated high-confidence subset for final local-control supervised fine-tuning. Third, we introduce training mechanisms that make zero-valued timing inputs numerically neutral and preserve the original behavior of the TTS backbone when no timing track is provided.

\begin{figure*}[!t]
\centering
\includegraphics[width=\textwidth]{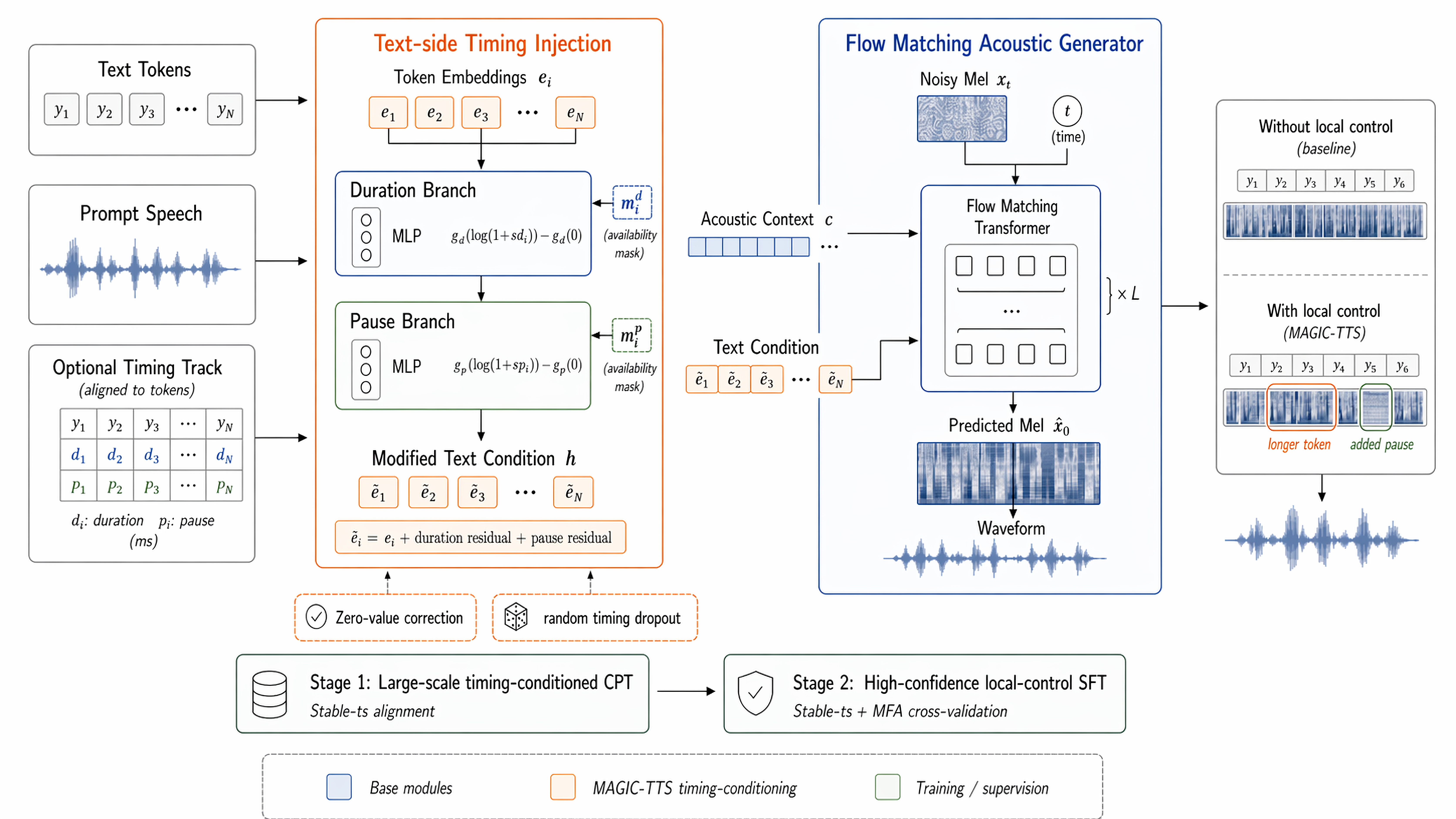}
\caption{Overview of MAGIC-TTS.}
\label{fig:method_overview}
\end{figure*}

Figure~\ref{fig:method_overview} summarizes how these components interact. The left panel defines the controllable inputs: the model always receives text tokens and prompt speech, and it optionally receives a token-aligned timing track specifying content duration and pause for each token. The middle panel shows the key architectural change of MAGIC-TTS, where the content-duration branch and pause branch encode numerical local controls, apply zero-value correction and availability masking, and inject the resulting residuals into the text-side condition. This design makes it possible to edit only selected token positions while leaving the rest of the utterance unconstrained. The right panel shows that the downstream acoustic generator remains the same flow-matching backbone, so the method adds local control by modifying the conditioning pathway rather than replacing the synthesis objective. The training-stage blocks at the bottom connect this architectural view to the data pipeline: Stage 1 uses large-scale timing-conditioned continued pretraining with Stable-ts labels to teach the model to use numerical timing cues, and Stage 2 uses high-confidence local-control supervised fine-tuning on the Stable-ts+MFA cross-validated subset to sharpen precise local controllability. In the current implementation, zero-value correction makes a zero timing input contribute no additional residual, while availability masking is primarily used to expose the model to all-zero timing conditions during training so that it can reliably recover spontaneous synthesis behavior when the timing track is absent.

\subsection{Flow-Based TTS with Local Timing Conditions}

We build MAGIC-TTS on a non-autoregressive zero-shot TTS backbone based on conditional flow matching. The model takes text tokens and an acoustic prompt as conditions, and generates the target mel-spectrogram through a parallel acoustic generator. This class of backbone is suitable for local timing control because the generation process is not a left-to-right rollout: timing conditions can be injected into the conditioning sequence and used by the acoustic generator when predicting all masked acoustic frames.

Let $\mathbf{x}_1$ be the target mel-spectrogram and $\mathbf{x}_0 \sim \mathcal{N}(0,I)$ be Gaussian noise with the same shape. During training, a time step $t \sim \mathcal{U}(0,1)$ is sampled and the noisy intermediate state is constructed by linear interpolation:
\begin{equation}
    \mathbf{x}_t = (1-t)\mathbf{x}_0 + t\mathbf{x}_1.
\end{equation}
The corresponding target flow is
\begin{equation}
    \mathbf{u}=\mathbf{x}_1-\mathbf{x}_0.
\end{equation}
The acoustic generator predicts a flow field conditioned on the acoustic context $\mathbf{c}$, the text-side conditioning sequence $\mathbf{h}$, and the diffusion time $t$:
\begin{equation}
    \hat{\mathbf{u}} = v_\theta(\mathbf{x}_t, t \mid \mathbf{c}, \mathbf{h}).
\end{equation}
Following the masked conditional generation setup of the backbone, the loss is applied to the target acoustic span:
\begin{equation}
    \mathcal{L}_{\mathrm{cfm}}
    = \mathrm{E}\left[\left\|
    \mathbf{M}\odot
    \left(v_\theta(\mathbf{x}_t,t\mid\mathbf{c},\mathbf{h})-\mathbf{u}\right)
    \right\|_2^2\right],
\end{equation}
where $\mathbf{M}$ denotes the acoustic mask. MAGIC-TTS keeps this objective unchanged. The key change is only in how the text condition $\mathbf{h}$ is formed.

For each token, we start from its text embedding $\mathbf{e}_i$ and add two timing-conditioned residuals, one for content duration and one for pause:
\begin{equation}
\begin{aligned}
    \tilde{\mathbf{e}}_i = \mathbf{e}_i
    &+ \alpha_d m_i^d \left(g_d(\log(1+s d_i)) - g_d(0)\right) \\
    &+ \alpha_p m_i^p \left(g_p(\log(1+s p_i)) - g_p(0)\right),
\end{aligned}
\end{equation}
where $g_d$ and $g_p$ are lightweight MLP encoders, $s$ is a log-scale factor, $m_i^d$ and $m_i^p$ indicate whether the corresponding controls are available, and $\alpha_d,\alpha_p$ are learnable gates. The logarithmic transform compresses the dynamic range of frame counts, so both short and long timing values can be represented smoothly. The learnable gates are initialized conservatively, allowing the model to start from the pretrained backbone behavior and gradually learn how much each timing branch should influence the text representation.

This residual design has two useful properties. First, it does not require a separate duration predictor or an autoregressive alignment module. The model directly receives the intended local timing values and learns to use them as conditions for acoustic generation. Second, it keeps the timing branch local and interpretable: $d_i$ controls the spoken duration of token $y_i$, while $p_i$ controls the pause associated with that token. Since the modified embeddings $\tilde{\mathbf{e}}_i$ are consumed by the same parallel flow-based acoustic generator, MAGIC-TTS can add explicit local timing control without replacing the synthesis backbone.

\subsection{Reliable Timing Supervision}

Reliable local control requires reliable local supervision. This requirement is stronger than ordinary duration modeling, because the model is not merely asked to predict plausible timing from text. It is asked to follow user-specified timing values at particular token positions. If the supervision assigns an incorrect boundary to a token, the same numerical value may correspond to different acoustic regions across samples. Such noise is especially harmful for content duration, since content control depends on accurate token interiors rather than only on whether a pause exists near a boundary.

MAGIC-TTS therefore prepares timing supervision in two complementary forms. For coverage, we first use Stable-ts alignments \citep{jianfch2025stablets} to construct token-level timing labels for the full continued-training corpus, which contains approximately 30k hours of speech. Word-level spans are projected onto the model's text tokens, and each token receives a pair $(d_i,p_i)$ corresponding to its content duration and pause in acoustic frames. This large-scale labeling stage exposes the pretrained flow-based backbone to explicit timing conditions across diverse speakers, texts, and acoustic environments. Its purpose is broad adaptation: the model learns that local numerical timing values can be used as generation conditions.

For precision, we further construct a high-confidence subset by cross-validating Stable-ts alignments with MFA alignments. The two alignment sources have different error patterns, so agreement between them provides a useful signal for filtering unreliable local timing labels. We first project both alignments onto a normalized text axis, where punctuation and formatting differences are removed and each aligned segment is associated with a span on the same canonical text sequence. We then retain an utterance only when three consistency checks are satisfied.

First, the two alignments must cover the same text range. This removes samples where one aligner skips words, inserts unmatched fragments, or aligns only a partial transcript. Second, the token groupings induced by the two alignments must be order-consistent. That is, after both alignments are mapped to token spans, their spans are not allowed to cross each other on the text axis. This criterion filters cases where one source merges or splits neighboring tokens in a way that would make token-level duration assignment ambiguous. Third, for matched spans that pass the text-side checks, their start and end times must be sufficiently close. This boundary-distance check removes samples where the two aligners agree on the words but disagree on the acoustic region assigned to them.

The cross-validated subset is intentionally conservative. Under this high-confidence setting, the retained data contains 202,086 utterances and 230.72 hours of speech. For these retained samples, we use the MFA alignment as the final timing source for local-control supervised fine-tuning. This stage provides cleaner supervision for precise timing following. In particular, the subset helps strengthen content-duration controllability: when token boundaries are reliable, increasing or decreasing $d_i$ has a consistent acoustic meaning, so the model can learn a more direct mapping from numerical content duration to local speech realization.

\subsection{Training for Balanced Practical Control}

MAGIC-TTS is trained so that local controls are useful when provided and harmless when absent. In practice, a TTS system may be used in an uncontrolled mode, where no timing track is supplied, or in a controlled mode, where content duration and pause are specified for every token. The training strategy therefore needs to satisfy two goals at the same time: it should make the timing track strong enough to control local speech, but it should not make the model depend on timing labels for ordinary synthesis.

\paragraph{Balancing content duration and pause.}
Content duration and pause are both represented as frame counts, but they do not create the same learning signal. Pause values are attached to every token position, and in natural fluent speech there is usually no pause between most neighboring tokens, so many of these values are zero. If a naive pause encoder maps zero to a nonzero vector, this vector is added densely across the text sequence even when no actual pause is requested. The pause branch can then become a strong global residual, making it easy for the model to rely on pause-related cues while reducing the relative effect of content-duration edits. This is undesirable because content duration is the more delicate control: it requires the model to stretch or compress the spoken realization of a specific token, not simply insert or remove silence near a boundary.

To avoid this imbalance, MAGIC-TTS uses zero-value correction in both timing branches by centering each timing encoder at its zero input. The timing residuals are computed as $g_d(\log(1+s d_i))-g_d(0)$ and $g_p(\log(1+s p_i))-g_p(0)$, so a true numerical zero contributes no timing residual before the learnable gate. This makes zero pause genuinely neutral, prevents the pause branch from introducing a dense bias over all token positions, and leaves a clearer learning signal for content-duration control. The same correction also makes the timing representation easier to interpret, because nonzero residuals correspond to actual nonzero timing conditions.

\paragraph{Robustness to missing controls.}
After zero-value correction, a true numerical zero contributes no additional timing residual; when a control is unavailable, the corresponding branch is also set to zero by the availability mask. The main purpose of availability masking is to expose the model to training examples in which the timing branch is entirely silent, so that all-zero residual patterns are learned as ordinary spontaneous synthesis conditions.

During training, we randomly drop the timing track for a subset of samples by setting the availability masks to zero. The model thus observes both controlled and uncontrolled cases under the same flow-matching objective. This missing-control training is important for preserving the default behavior of the flow-based backbone: the model cannot assume that timing labels will always be present, and it must still synthesize fluent speech from text and prompt alone. At inference time, users can provide a full timing track or omit the track entirely. This allows MAGIC-TTS to support fine-grained local timing manipulation while remaining compatible with ordinary zero-shot TTS usage.

\section{Experiments}
\begin{table*}[!t]
\centering
\small
\setlength{\tabcolsep}{4.5pt}
\begin{tabular}{lcccccc}
\toprule
Setting & C-MAE $\downarrow$ & P-MAE $\downarrow$ & C-Corr. $\uparrow$ & P-Corr. $\uparrow$ & F1@50 $\uparrow$ & F1@100 $\uparrow$ \\
\midrule
F5-TTS Base & 38.82 & 20.68 & 0.594 & 0.225 & 0.129 & 0.118 \\
Baseline CPT final & 40.65 & 19.10 & 0.562 & 0.293 & 0.149 & 0.144 \\
MAGIC CPT final (controlled) & 15.93 & 10.45 & 0.787 & 0.734 & 0.405 & 0.400 \\
MAGIC CPT final (spontaneous) & 38.41 & 21.37 & 0.599 & 0.260 & 0.137 & 0.133 \\
SFT w/ timing control (controlled) & \textbf{10.56} & \textbf{8.32} & \textbf{0.918} & \textbf{0.793} & \textbf{0.410} & \textbf{0.397} \\
SFT w/ timing control (spontaneous) & 36.88 & 18.92 & 0.588 & 0.283 & 0.128 & 0.113 \\
\bottomrule
\end{tabular}
\caption{Timing control accuracy comparison. The table is organized to separate control-free baselines, CPT-final checkpoints with local-timing branches, and the final SFT model under both controlled and spontaneous inference modes.}
\label{tab:reference_timing}
\end{table*}

\begin{table}[!t]
\centering
\footnotesize
\setlength{\tabcolsep}{3.5pt}
\resizebox{\columnwidth}{!}{%
\begin{tabular}{lccccc}
\toprule
Type & Base Tgt. & Base Mean & Edit Tgt. & Edit Mean & Abs. Bias $\downarrow$ \\
\midrule
Content & 170.00 & 171.07 & 225.00 & 207.40 & 17.60 \\
Pause & 0.00 & 0.00 & 260.00 & 236.67 & 23.33 \\
\bottomrule
\end{tabular}
}
\caption{Scenario-based local timing editing benchmark averaged over the three retained demos.}
\label{tab:local_editing}
\end{table}

\begin{table*}[!t]
\centering
\small
\setlength{\tabcolsep}{4.0pt}
\resizebox{\textwidth}{!}{%
\begin{tabular}{lcccccc}
\toprule
System & C-MAE $\downarrow$ & P-MAE $\downarrow$ & C-Corr. $\uparrow$ & P-Corr. $\uparrow$ & F1@50 $\uparrow$ & F1@100 $\uparrow$ \\
\midrule
MAGIC-TTS & 10.56 & 8.32 & 0.918 & 0.793 & 0.410 & 0.397 \\
w/o zero correction & 12.89 & 9.48 & 0.890 & 0.793 & 0.428 & 0.388 \\
w/o cross-validated timing supervision & 15.93 & 10.45 & 0.787 & 0.734 & 0.405 & 0.400 \\
\bottomrule
\end{tabular}
}
\caption{Controllability ablation of MAGIC-TTS under controlled synthesis.}
\label{tab:ablation}
\end{table*}

\subsection{Experimental Setup}

\paragraph{Model and training.}
We instantiate MAGIC-TTS with the official F5-TTS Base configuration as the zero-shot TTS backbone \citep{chen2024f5tts}. Concretely, the acoustic generator is a DiT-based conditional flow-matching model with hidden size 1024, 22 transformer blocks, 16 attention heads, feed-forward multiplier 2, text-conditioning dimension 512, and 4 text-side convolution layers. The model predicts 100-bin mel-spectrograms at 24 kHz and uses a Vocos-based acoustic representation and tokenizer setting. The learnable content and pause gates are initialized to 0. All training runs are conducted on a single node with 8 NVIDIA A800 GPUs and 64 CPU cores. Unless otherwise specified, MAGIC-TTS is evaluated in two modes: \emph{spontaneous}, where no timing track is provided, and \emph{controlled}, where token-level content duration and pause values are supplied.

For duration-conditioned continued training, we start from the official F5-TTS Base checkpoint and use a large Emilia subset whose transcripts are re-decoded by the MNV-17 NV-aware ASR model \citep{mai2025mnv17}. We retain samples predicted to contain at least one nonverbal vocalization, so that the continued-training corpus preserves expressive and nonverbal-rich speaking styles. This also makes the training set rich in sequence patterns where semantic content and pauses are interleaved, forcing the model to rely more on the provided local duration prior when generating speech. The prepared flow dataset contains 2,195,557 utterances with Stable-ts-derived token-level timing labels \citep{jianfch2025stablets}. We train with dynamic batching of 30,000 audio frames per GPU, gradient accumulation 1, maximum gradient norm 1.0, learning rate $7.5\times 10^{-5}$, warmup for 20,000 updates, and duration-dropout probability 0.2. This stage is run for 2 epochs and reaches 27,000 updates in total.

For high-confidence local-control fine-tuning, we use the cross-validated B@150 subset described in Section~\ref{sec:method}. The final reported checkpoint is taken at SFT step 36,000. This stage uses dynamic batching of 30,000 audio frames per GPU, gradient accumulation 1, maximum gradient norm 1.0, learning rate $7.5\times 10^{-5}$, warmup for 1,000 updates, and duration-dropout probability 0.2.

\paragraph{Cross-validation configuration.}
The high-confidence subset is constructed with the B@150 filter. We first project Stable-ts \citep{jianfch2025stablets} and MFA word alignments onto a shared normalized text axis by lowercasing the text and retaining only alphanumeric characters and Chinese characters, which removes punctuation and formatting differences. We then compare the two alignment sequences on the normalized axis. An utterance is retained only when three conditions are simultaneously satisfied: 1) the two alignments cover the same text range, 2) their induced token groupings remain order-consistent and do not create crossing boundaries on the text axis, and 3) for every comparable local span, the maximum of start-time difference, end-time difference, and span-duration difference is no greater than 150 ms. Starting from 13,627,216 EN/ZH entries, 13,062,413 have both alignments available for comparison, 202,164 pass the B@150 filter, and 202,086 remain after dataset materialization cleanup, corresponding to 230.72 hours of speech.

\paragraph{Evaluation protocol.}
We evaluate MAGIC-TTS from two perspectives. First, we test whether the model obeys explicit numerical local timing controls when token-level content duration and pause values are provided. Second, we test whether local edits remain localized. For timing evaluation, we synthesize speech from text and prompt speech, force-align the generated waveform with MFA, and compare the realized token-level content duration and pause with the prescribed target values.

\paragraph{Baselines and ablations.}
For method analysis, we compare with ablated variants of MAGIC-TTS. The main controllability ablations remove the major design choices that are intended to directly affect explicit timing control: zero-value correction and cross-validated timing supervision.

\subsection{Timing Control Accuracy}

We next evaluate whether MAGIC-TTS can realize prescribed token-level timing controls. We use 100 utterances randomly sampled from the B@150 subset with durations between 3 and 10 seconds as a test set. For each test sample, the target token-level content duration and pause values are extracted from the ground-truth MFA alignment. We then synthesize the same text in two modes. In the spontaneous mode, the model receives no local timing controls. In the controlled mode, the model receives the target token-level content-duration and pause values. Unless otherwise specified, controlled inference uses the full prompt+target text together with the full prompt+target timing track, which matches the training-time conditioning format. After synthesis, we force-align the generated speech with MFA and compare the realized timing with the prescribed target values.

This evaluation directly measures whether explicit token-level controls take effect. If the controls are effective, controlled synthesis should produce content durations and pauses that are closer to the prescribed target values than spontaneous synthesis. We report mean absolute error (MAE) for content duration and pause, as well as correlation between target and realized timing values. For pause placement, we additionally report a threshold-based F1 score, where a boundary is treated as a pause when its duration exceeds a fixed threshold.

Table~\ref{tab:reference_timing} shows a clear and consistent advantage for explicit local timing control. Even in the spontaneous condition, the model still exhibits a moderate correlation with the target timing, which is expected because natural prosody and segmental duration are partly predictable from text and prompt speech. However, once explicit token-level content duration and pause values are provided, all timing metrics improve sharply. Content-duration MAE decreases from 36.88 ms to 10.56 ms, while content correlation increases from 0.588 to 0.918. This result is especially important because content-duration control is the more delicate part of the problem: the model must reshape the spoken realization of a token itself, rather than merely insert silence at a boundary. The magnitude of the gain indicates that MAGIC-TTS does not treat the duration track as a weak auxiliary hint; it learns to use the numerical token-local conditions as a primary signal for allocating acoustic time within the utterance.

Pause control also improves strongly and consistently. Pause MAE drops from 18.92 ms to 8.32 ms, pause correlation increases from 0.283 to 0.793, and pause F1 improves from 0.128 to 0.410 under the 50 ms threshold and from 0.113 to 0.397 under the 100 ms threshold. These gains show that explicit pause conditions improve not only the amount of pause realized by the model, but also whether a pause is placed at the correct boundary at all. This behavior matches the design intention of MAGIC-TTS: pause is represented as an explicit token-local variable and injected directly into the flow-based acoustic generator, so the model can respond to local boundary timing instructions in a stable and interpretable way.

Taken together, these results strongly support the effectiveness of our method. First, they verify that the explicit timing-conditioning mechanism is actually used at inference time, rather than being ignored by a strong pretrained backbone. Second, they show that the gains are not limited to a single easy metric: MAGIC-TTS improves absolute error, correlation, and threshold-based pause-detection consistency at the same time. Third, the relative pattern of improvement is itself informative and favorable. The especially large content-duration gain indicates that the model can use the explicit track to regulate token-internal timing, while the strong pause gains show that the same mechanism also provides reliable leverage over boundary timing. This combination is precisely what we want from a practical local-control TTS system: strong pause controllability, meaningful content-duration controllability, and consistent behavior across multiple evaluation views.

We also note that the evaluation is conservative in two ways. Only successfully aligned samples are scored, leaving 92 controlled cases and 90 spontaneous cases after filtering. These alignment failures come from MFA itself rather than obvious synthesis collapse, since force alignment occasionally fails on otherwise usable generated samples. Moreover, the target timing is compared against timing re-extracted from generated speech via MFA, so the reported values are additionally affected by MFA variability and boundary jitter.

\subsection{Local Timing Editing}

Beyond full-track following, we also test whether MAGIC-TTS supports practical local editing scenarios. We build a small scenario-based editing benchmark with three retained demos. The benchmark text is written in Chinese, and we provide English glosses together with pinyin references: navigation guidance (\emph{qian fang lu kou zuo zhuan}; \emph{``At the next intersection, turn left.''}), guided reading (\emph{gen wo du, ping guo}; \emph{``Read after me, apple.''}), and accessibility-oriented code reading (\emph{yan zheng ma shi 379, 218}; \emph{``The verification code is 379, 218.''}). In all three cases, synthesis starts from a \emph{uniform-timing baseline} track in which content tokens are assigned 170\,ms and punctuation is assigned 50\,ms. We then modify only the selected pause location and the selected content tokens, while leaving the remaining track unchanged.

This benchmark serves two purposes. First, it verifies precise duration control under a uniform baseline track: before any local edit, the model should already realize the prescribed per-token baseline with small bias. Second, it checks edit effectiveness: once only a few local entries are changed, the edited realization should move toward the requested local targets. Because the three scenarios contain different numbers of edited tokens, we summarize the benchmark by reporting mean baseline realization, mean edited realization, and the absolute bias between the mean edited realization and the mean target.

Table~\ref{tab:local_editing} shows that the benchmark starts from a reproducible baseline.

After local edits are applied, the model moves the targeted regions substantially toward the requested values. Across the three retained demos, the edited content mean reaches 207.40\,ms for a 225\,ms target, with 17.60\,ms absolute bias. The edited pause mean reaches 236.67\,ms for a 260\,ms target, with 23.33\,ms absolute bias.

At the same time, the MFA-based measurements in this benchmark are not perfectly exact, especially for short local spans. Under this relatively low baseline bias, however, we still view the benchmark as evidence for practical local editing. The result shows that MAGIC-TTS can faithfully realize a uniform timing track at the token level, and can also precisely lengthen selected tokens or insert pauses when only a few local entries are changed.

We further evaluate larger-scale local editing on B@150 with both isolated single-token edits and adjacent multi-token span edits. Because these stress tests require more careful treatment of Chinese token reconstruction from MFA word boundaries, and because we additionally report language-split tables under the strict content-evaluation rule, we move the full protocol, six detailed tables, and the corresponding analysis to Appendix~\ref{sec:appendix_local_editing}. The appendix confirms the same qualitative conclusion as the small scenario benchmark: isolated single-token edits are intentionally harsh, while adjacent short-span edits are more practical and substantially stronger, especially in the stretch direction.

\subsection{Ablation Studies}

Finally, we ablate the main components of MAGIC-TTS that are intended to directly improve explicit timing controllability. Removing zero-value correction tests whether zero duration and zero pause values introduce harmful residual bias. This ablation is especially relevant for balancing content duration and pause, because pause values occur at every token position and many of them are zero. Removing cross-validated timing supervision tests whether cleaner token-level timing labels are needed for precise local control, especially for content duration. We restrict Table~\ref{tab:ablation} to controllability metrics only.

Table~\ref{tab:ablation} should be read with the asymmetry between content duration and pause in mind. In general, pause is the easier variable for the model to exploit: before our control-oriented refinements, the model can already rely heavily on pause duration because pause is attached to every token position and is easier to realize acoustically than token-internal stretching or compression. As a result, ablated systems can obtain slightly higher pause-side correlation or thresholded pause F1 by overusing pause-like behavior, even though they remain weaker at the more difficult and practically more important task of precise content-duration control.

From this perspective, the relative pattern in Table~\ref{tab:ablation} is consistent with our design goal. Zero-value correction suppresses the dense residual bias introduced by frequent zero-pause entries, which reduces the model's tendency to over-rely on pause conditioning. Cross-validated timing supervision further gives the model cleaner token-boundary information, strengthening its ability to interpret and follow content-duration targets precisely. MAGIC-TTS therefore improves the content-side metrics most clearly, while keeping pause control strong overall rather than maximizing the easiest pause-oriented metrics at the expense of balanced local controllability.

\section{Conclusion}

We presented MAGIC-TTS, a flow-based zero-shot TTS system with explicit token-level control over both content duration and pause. By injecting numerical local timing conditions and training with zero-value correction, missing-control robustness, and cross-validated timing supervision, MAGIC-TTS achieves strong controllability on both token-internal timing and boundary timing. The experiments further show that this control interface supports practical local editing, while introducing only a moderate trade-off in uncontrolled generation quality for the intended controllable-synthesis setting.

\bibliography{refs}

\clearpage
\appendix
\section{B@150 Local Editing Stress Tests}
\label{sec:appendix_local_editing}

We report the full B@150 local-editing stress tests in this appendix. These experiments complement the three-demo scenario benchmark in the main text by probing stronger and more localized edits on a larger evaluation set. We consider two settings: a \emph{single-token stress test}, where one editable target-side token is modified in the generated half of each utterance, and an \emph{adjacent multi-token editing} setup, where a short two-token or three-token region is edited jointly. For the single-token setting, we retain only pause targets of 500\,ms and 800\,ms as formal quantitative results. We did also run single-token content edits with factors 0.5$\times$, 1.5$\times$, and 2.0$\times$, but manual listening evaluation showed that word-level MFA boundaries are too unstable for this setting, so the mean values of these content-side measurements are less informative than their distributional behavior. We further found that even under the strict boundary filter, the 200\,ms pause condition is still too fragile for word-level MFA evaluation: manual listening confirms cases where an audible pause is inserted, but MFA does not expose that boundary as an explicit gap, producing false zero measurements. We therefore do not include the 200\,ms condition in the formal quantitative evaluation set. For the multi-token setting, we evaluate the same three duration factors on both two-token and three-token regions, and the realized edit span is measured from the first edited token onset to the last edited token offset, including internal pauses.

For editing evaluation, the reported aggregates use a strict rule. For content edits, we ensure that the two edit boundaries do not fall inside the content of an MFA word. For pause edits, we ensure that the edited boundary itself does not fall inside the content of an MFA word.

\begin{figure*}[!t]
\centering
\includegraphics[width=0.94\textwidth]{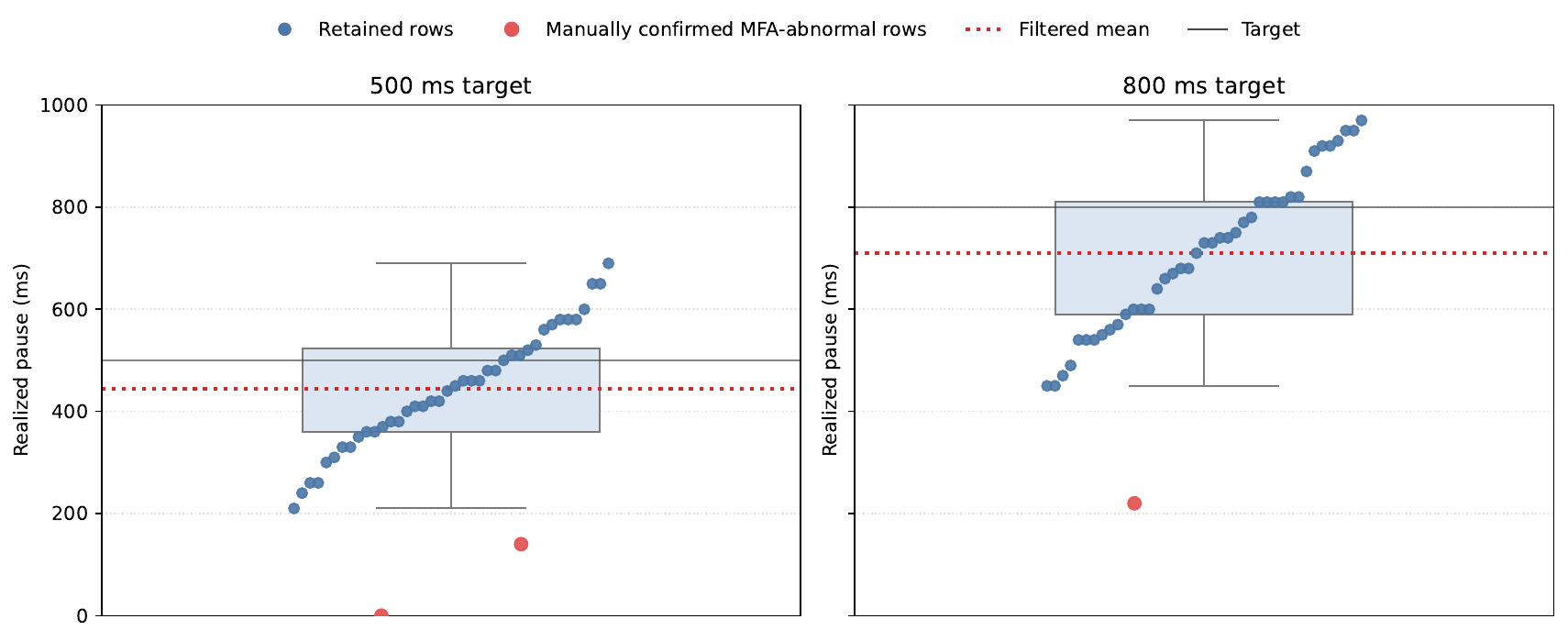}
\caption{Single-token pause-only local editing in the B150 test set. Blue points denote the retained rows after removing manually confirmed MFA-abnormal cases, red points denote the removed abnormal cases, the dotted line is the filtered mean, and the solid horizontal line is the target pause.}
\label{fig:appendix_single_token_pause_distribution}
\end{figure*}

Instead of reporting a mean-only table, Figure~\ref{fig:appendix_single_token_pause_distribution} reports the full distribution, because the measurement tool itself is not fully reliable. For content edits, manual listening evaluation shows that MFA can shift short material across neighboring word boundaries, which makes token-local realized-duration means less informative than the overall distribution. For pause edits, MFA tends to absorb the left and right pause boundaries into adjacent content regions, so the effective pause is often underestimated. As a result, the whole distribution is somewhat biased low, not only the extreme outliers.

We therefore report the realized-pause distribution directly and treat the mean only as a secondary summary. In the strict set, the raw means are 426.19\,ms for the 500\,ms condition and 698.81\,ms for the 800\,ms condition. After removing the manually confirmed MFA-abnormal rows highlighted in red, the means increase to 444.00\,ms and 710.49\,ms, respectively. These filtered values are still conservative, because even many non-red points remain subject to the same under-measurement tendency. Therefore, the residual gap to target should not be read as quantitative evidence that MAGIC-TTS follows pause instructions imprecisely; instead, it provides a conservative reference for the model's achievable pause-control range.

The most concrete failure mode we observed is rare but informative. Across the 84 strict-boundary rows in the retained 500\,ms and 800\,ms single-token pause set, we found one utterance whose pause mass was clearly attracted to a nearby pre-existing pause instead of remaining centered on the requested boundary. The utterance is the Chinese sentence transliterated as ``lian xuexiao de yundongfu hai mei laideji huan jiu bei muqin misha''. In the baseline realization, there is already an audible pause at the boundary ``huan | jiu'', while the requested edit asks for a new pause at ``jiu | bei''. For this utterance, MAGIC-TTS partially or fully absorbs the requested pause into the earlier ``huan | jiu'' interruption: in the 500\,ms condition, the previous pause expands from about 200\,ms to about 500\,ms while the target boundary after ``jiu'' remains at 0\,ms under MFA; in the 800\,ms condition, the previous pause also grows while the target boundary is exposed only partially. We treat this as a concrete failure mode of pause-location fidelity. A plausible explanation is that this inference pattern is unusually far from the duration configurations seen during training, so the model chooses a nearby acoustically easier boundary and ``attracts'' the requested pause mass there. Importantly, this behavior was found in only one utterance, so it is a real but low-probability failure pattern rather than the dominant behavior of the model.

\begin{table*}[!t]
\centering
\small
\setlength{\tabcolsep}{4.5pt}
\begin{tabular}{lcccc}
\toprule
Span & Target & N & Realized & Error / Neighbor Drift \\
\midrule
2-token & 0.5$\times$ & 58 & 0.685$\times$ & 78.02 ms / 14.93 ms \\
2-token & 1.5$\times$ & 58 & 1.378$\times$ & 83.19 ms / 10.36 ms \\
2-token & 2.0$\times$ & 58 & 1.777$\times$ & 132.93 ms / 18.58 ms \\
3-token & 0.5$\times$ & 55 & 0.666$\times$ & 115.48 ms / 16.70 ms \\
3-token & 1.5$\times$ & 55 & 1.388$\times$ & 112.24 ms / 14.26 ms \\
3-token & 2.0$\times$ & 55 & 1.783$\times$ & 207.66 ms / 17.13 ms \\
\bottomrule
\end{tabular}
\caption{Adjacent multi-token local editing on B@150, aggregated under the strict content-evaluation rule. Realized values are measured on the full edited span from the first edited token onset to the last edited token offset.}
\label{tab:appendix_multitoken_overall}
\end{table*}

\begin{table*}[!t]
\centering
\small
\setlength{\tabcolsep}{4.5pt}
\begin{tabular}{lcccc}
\toprule
Span & Target & N & Realized & Error / Neighbor Drift \\
\midrule
2-token & 0.5$\times$ & 40 & 0.657$\times$ & 78.37 ms / 9.12 ms \\
2-token & 1.5$\times$ & 40 & 1.395$\times$ & 81.38 ms / 7.75 ms \\
2-token & 2.0$\times$ & 40 & 1.806$\times$ & 126.25 ms / 9.13 ms \\
3-token & 0.5$\times$ & 40 & 0.641$\times$ & 106.65 ms / 10.52 ms \\
3-token & 1.5$\times$ & 40 & 1.410$\times$ & 108.95 ms / 9.02 ms \\
3-token & 2.0$\times$ & 40 & 1.833$\times$ & 202.78 ms / 9.78 ms \\
\bottomrule
\end{tabular}
\caption{English-only results for adjacent multi-token local editing on B@150.}
\label{tab:appendix_multitoken_en}
\end{table*}

\begin{table*}[!t]
\centering
\small
\setlength{\tabcolsep}{4.5pt}
\begin{tabular}{lcccc}
\toprule
Span & Target & N & Realized & Error / Neighbor Drift \\
\midrule
2-token & 0.5$\times$ & 18 & 0.745$\times$ & 77.22 ms / 27.85 ms \\
2-token & 1.5$\times$ & 18 & 1.340$\times$ & 87.22 ms / 16.18 ms \\
2-token & 2.0$\times$ & 18 & 1.710$\times$ & 147.78 ms / 39.56 ms \\
3-token & 0.5$\times$ & 14 & 0.736$\times$ & 135.36 ms / 34.32 ms \\
3-token & 1.5$\times$ & 14 & 1.336$\times$ & 108.21 ms / 28.65 ms \\
3-token & 2.0$\times$ & 14 & 1.649$\times$ & 211.43 ms / 36.67 ms \\
\bottomrule
\end{tabular}
\caption{Chinese-only results for adjacent multi-token local editing on B@150.}
\label{tab:appendix_multitoken_zh}
\end{table*}

\begin{figure*}[!t]
\centering
\includegraphics[width=0.98\linewidth]{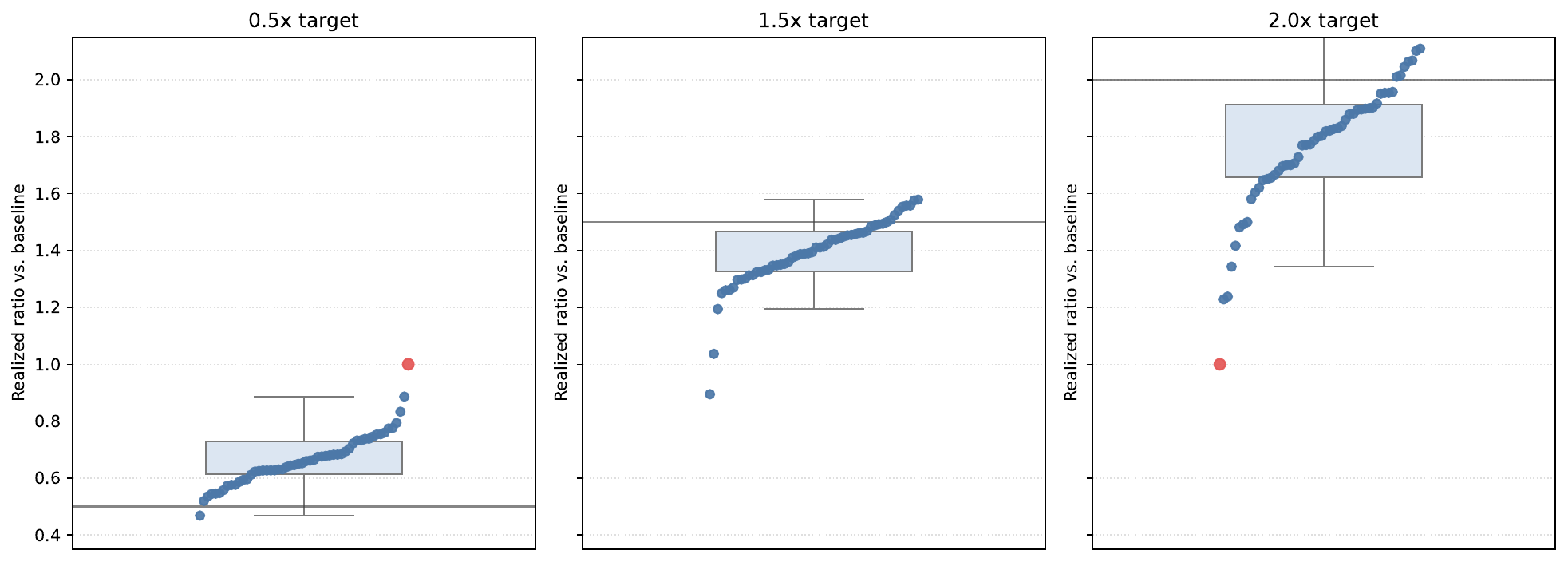}
\caption{Distribution view for the mixed 3-token content-editing set on B@150. Each panel corresponds to one requested factor, the horizontal solid line marks the target factor, and red points denote listening-confirmed cases where MFA conservatively mismeasures the realized edit because part of the edited acoustic span is reassigned to a neighboring word.}
\label{fig:appendix_multitoken_zh_dist}
\end{figure*}

Tables~\ref{tab:appendix_multitoken_overall}--\ref{tab:appendix_multitoken_zh} show that adjacent-span editing is considerably more practical than the isolated single-token stress test. In the strict aggregate, the 2.0$\times$ condition reaches 1.777$\times$ for two-token spans and 1.783$\times$ for three-token spans, clearly stronger than the corresponding single-token result. The English subset remains the strongest and most stable, with 1.806$\times$ and 1.833$\times$ realized ratios at 2.0$\times$ for two-token and three-token spans, respectively. The Chinese strict subset is again smaller, retaining only 18 of 48 Chinese rows for two-token spans and 14 of 48 for three-token spans under the text-boundary rule.

At the same time, we do not interpret these editing tables as exact acoustic truth. Manual listening evaluation shows that MFA can conservatively underestimate both compression and stretch by reassigning part of the edited acoustic span into a neighboring word. In the compression direction, a sample from `chen2 guang1 jiu4 yi4 zhi2 mo4 ming2 qi2 miao4 de shuo1 shen2 me nv3 peng2 you3' audibly compresses the edited region `shuo1 shen2 me', but MFA absorbs part of the following `nv3 peng2 you3' onset into the edited region and therefore reports an overly weak realized ratio. In the stretch direction, a sample from `dan4 shi4 ta1 de xing4 neng2 que4 yuan3 yuan3 chao1 guo4 le qi2 ta1 shang1 pin3' audibly lengthens the edited region `le qi2 ta1', but MFA assigns part of `qi2 ta1' to the following word `shang1 pin3' and therefore reports a realized ratio that is too small. Figure~\ref{fig:appendix_multitoken_zh_dist} makes this conservative measurement pattern visible directly. We therefore view all reported editing-test tables as conservative lower-bound summaries rather than as fully faithful measurements of the acoustic edit strength.

Finally, we note that the strict evaluation rule is still useful even though it does not eliminate every MFA artifact. We constructed MFA-sliced listening packs for representative English and Chinese local-editing samples, and manually checked both pause-only edits and content edits. These manual checks show that the reported editing measurements remain informative, but they are better interpreted as conservative lower-bound summaries than as exact acoustic truth in every case.

\section{Additional Analysis}
\label{sec:appendix_content_dynamics}

{\bf Why explicit token duration in prior TTS does not automatically imply a MAGIC-style control interface.}
During internal probing, we also tested a zero-training duration-injection baseline on an external duration-based TTS system, Grad-TTS. The purpose was not to introduce a new formal benchmark, but to clarify a conceptual point in the related-work discussion: even when a prior TTS system contains explicit token-level duration variables, those variables do not necessarily function as reliable user-facing local controls. In our probe, manually replacing Grad-TTS's internal duration expansion with MFA-derived target-side durations did shift the generated rhythm scaffold toward the target, but both timbre quality and semantic clarity degraded sharply: the synthesized speech became markedly worse in voice fidelity, and the spoken content itself became much less clear. This behavior supports an important distinction. In MAGIC-TTS, duration is trained as a high-confidence external control prior: the model is explicitly exposed to numerical local timing conditions and gradually learns to rely on them when deciding token-local boundary placement and acoustic time allocation. In Grad-TTS, by contrast, duration mainly serves as an internal alignment/expansion intermediate predicted by the model itself. It helps organize the generator's timing scaffold, but it is not optimized as a strong user-facing control interface. As a result, manually replacing that intermediate can alter the rhythm skeleton without preserving the generator's learned coupling between duration, alignment, and acoustic realization. This difference is further amplified by pause modeling: MAGIC-TTS separately conditions on content duration and pause, whereas traditional duration-based systems typically absorb boundary pauses into the same duration variable. We therefore view prior explicit-duration TTS systems as evidence that token-duration information is useful, but not as evidence that a dependable local timing-control interface already exists.

We then extended the same zero-training probe to YourTTS, a zero-shot speaker-cloning VITS system. This second probe was informative for a different reason. YourTTS exposes an internal duration path at inference and also inserts explicit blank tokens between visible text characters. In principle, this makes it easier than Grad-TTS to inject pause-like timing by assigning external gap durations to blank positions. In practice, however, listening tests again showed that manually replacing these internal timing variables did not yield a clean user-facing control interface. Compared with Grad-TTS, YourTTS could more easily absorb externally specified pauses, but the resulting speech often became simultaneously much blurrier in timbre and much less intelligible in content, while the subjective rhythm shift toward the target was even more limited and was often barely perceptible in listening. This suggests that the problem is not only the absence of a separate pause variable. Even when a system exposes both visible-token duration and blank-token timing, those variables can still function primarily as internal coordination scaffolds shared by the duration predictor, latent flow prior, speaker conditioning path, and waveform decoder, rather than as semantically clean external controls. Our additional pause-aware blank probe reinforces this point: by assigning zero duration to blank tokens without MFA gaps and assigning positive duration only to blank positions that follow observed inter-word gaps, we could force the total duration to numerically match the target span, yet listening quality still degraded and target-like rhythm transfer remained weak. We therefore interpret the YourTTS probe as complementary evidence for the same core claim: exposing internal explicit timing variables is not sufficient by itself to create a strong local timing-control interface. What matters is whether the acoustic generator has been trained to rely on a high-confidence external duration-and-pause prior as part of its normal generation behavior.

Finally, we note a practical boundary of this probing line. After systems such as Grad-TTS and YourTTS, the mainstream zero-shot TTS trajectory increasingly shifted toward architectures that do not rely on directly exposed internal explicit token-duration modeling in the same way, or that entangle timing with broader latent-sequence generation mechanisms. For those later systems, the present ``manually replace internal token duration'' probe becomes less comparable and less diagnostically clean. We therefore stop this auxiliary baseline line at Grad-TTS and YourTTS. This boundary is not meant to imply that later systems lack any timing structure, but rather that they no longer provide the same kind of explicit internal duration variable needed for a like-for-like zero-training intervention.

\begin{table*}[!t]
\centering
\small
\setlength{\tabcolsep}{4.5pt}
\begin{tabular}{lcccc}
\toprule
System & EN WER $\downarrow$ & EN SIM $\uparrow$ & ZH CER $\downarrow$ & ZH SIM $\uparrow$ \\
\midrule
GT & 2.160 & 0.734 & 1.254 & 0.755 \\
VOC & 2.164 & 0.697 & 1.276 & 0.720 \\
F5 base & 1.993 & 0.667 & 1.665 & 0.744 \\
Baseline CPT final & 1.994 & 0.649 & 1.772 & 0.733 \\
MAGIC CPT final & 2.521 & 0.646 & 2.322 & 0.731 \\
MAGIC SFT final & 3.434 & 0.638 & 2.215 & 0.738 \\
\bottomrule
\end{tabular}
\caption{Additional uncontrolled quality comparison on Seed-TTS-Eval. `VOC' denotes the reference baseline obtained by reconstructing the ground-truth audio through the vocoder path, so it isolates the degradation introduced by the vocoder itself. `Baseline CPT final' denotes a fairness-oriented control-free continued-training baseline trained on the same large-scale data regime, `MAGIC CPT final' denotes the key controllability-focused checkpoint at the end of CPT, and `MAGIC SFT final' denotes the final controllable model evaluated in spontaneous mode without any timing track.}
\label{tab:appendix_uncontrolled_fairness}
\end{table*}

{\bf Trade-off Between Controllability and Generation Quality.}
To better isolate the effect of explicit controllability from the effect of large-scale continued training itself, we additionally trained a control-free baseline on the same large-scale data regime and evaluated it on Seed-TTS-Eval. Table~\ref{tab:appendix_uncontrolled_fairness} reports the reference systems (`GT', `VOC', and `F5 base'), the matched control-free continued-training baseline (`Baseline CPT final'), the key controllability-focused checkpoint at the end of CPT (`MAGIC CPT final'), and the final controllable model evaluated in spontaneous mode (`MAGIC SFT final'). Here `VOC' is included specifically to factor out vocoder influence: it reconstructs the ground-truth audio through the vocoder path and therefore provides the quality ceiling after vocoder-only degradation. This comparison is intended as a fairness-oriented supplement: if spontaneous quality drops slightly after adding local timing control, we want to separate the contribution of the control mechanism from the contribution of the matched continued-training setup.

The results show that uncontrolled quality does exhibit a mild regression after introducing explicit local control, but the magnitude remains limited. Relative to the matched control-free baseline, `MAGIC SFT final' changes from 1.994 to 3.434 on English WER and from 1.772 to 2.215 on Chinese CER, while speaker similarity stays close, changing from 0.649 to 0.638 in English and from 0.733 to 0.738 in Chinese. The `MAGIC CPT final' checkpoint is also informative: it stays noticeably closer to the control-free baseline than the final SFT model, with 2.521 English WER, 2.322 Chinese CER, 0.646 English SIM, and 0.731 Chinese SIM. This pattern supports the interpretation that the spontaneous-quality cost is introduced gradually during controllability-oriented training rather than being an unavoidable consequence of large-scale continued training alone. We therefore view the quality cost as being within a controllable range rather than indicating a failure of the backbone. More importantly, the main purpose of MAGIC-TTS is not to maximize spontaneous quality at any cost, but to add explicit token-local timing control that does not exist in the baseline system. Under this perspective, a modest uncontrolled-quality trade-off is acceptable, especially because the model gains strong and directly measurable controllability over local duration and pause, which is the central contribution of this work.

\begin{table*}[!t]
\centering
\footnotesize
\setlength{\tabcolsep}{3.5pt}
\begin{tabular}{lcccccc}
\toprule
Fmt. & C-MAE $\downarrow$ & P-MAE $\downarrow$ & C-Corr. $\uparrow$ & P-Corr. $\uparrow$ & F1@50 $\uparrow$ & F1@100 $\uparrow$ \\
\midrule
T-only & 27.98 & 17.34 & 0.659 & 0.462 & 0.279 & 0.272 \\
PM-free & 23.58 & 17.00 & 0.773 & 0.543 & 0.356 & 0.330 \\
Full cond. & \textbf{11.85} & \textbf{9.00} & \textbf{0.916} & \textbf{0.769} & \textbf{0.413} & \textbf{0.359} \\
\bottomrule
\end{tabular}
\caption{Inference-format ablation for timing control on the 100-sample B@150 evaluation set. Removing prompt-side duration conditions weakens controllability, and a model trained with prompt-side duration masking only partially recovers this gap.}
\label{tab:appendix_inference_format}
\end{table*}

{\bf Inference-Format Ablation.}
We report an inference-format ablation on an earlier MAGIC-TTS checkpoint trained for 20k SFT steps. Throughout this appendix, all reported training-step counts refer to \emph{SFT steps only}; they do not include the earlier continued-pretraining (CPT) stage, and SFT counting starts after the 27k-step CPT stage has finished. In Table~\ref{tab:appendix_inference_format}, `T-only` denotes target-only duration conditioning at inference with prompt-side duration removed, `PM-free` denotes the prompt-duration-masked model evaluated under the same prompt-duration-free setting, and `Full cond.` denotes inference with full prompt+target text and full prompt-side plus target-side duration conditioning. When prompt-side duration conditions are removed at inference and timing is provided only on the target side, timing control becomes noticeably weaker: content MAE rises from 11.85\,ms to 27.98\,ms, pause MAE rises from 9.00\,ms to 17.34\,ms, content correlation drops from 0.916 to 0.659, and pause correlation drops from 0.769 to 0.462. We further tested a prompt-duration-masked MAGIC-TTS variant under this same prompt-duration-free setting. Its best in-domain checkpoint reaches content MAE 23.58\,ms, pause MAE 17.00\,ms, content correlation 0.773, pause correlation 0.543, F1@50 0.356, and F1@100 0.330. This gives only a mild improvement over the corresponding setting without prompt-time-mask training, and it still remains clearly below the full-conditioning setup. The result suggests that prompt-time-mask training can reduce the model's dependence on prompt-side timing to a limited extent, but the strongest controllability still comes from retaining the full prompt+target timing format.

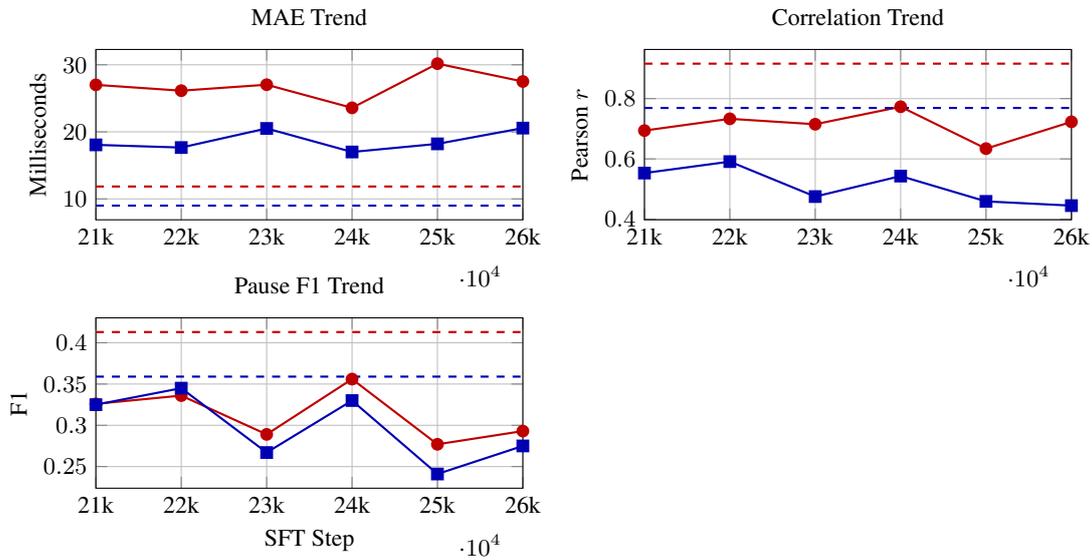
\begin{figure*}[!t]
\centering
\footnotesize
\begin{tikzpicture}
\begin{groupplot}[
    group style={group size=2 by 2, horizontal sep=1.6cm, vertical sep=1.3cm},
    width=0.45\textwidth,
    height=0.24\textwidth,
    grid=major,
    xmin=21000, xmax=26000,
    xtick={21000,22000,23000,24000,25000,26000},
    xticklabels={21k,22k,23k,24k,25k,26k},
    tick label style={font=\footnotesize},
    label style={font=\footnotesize},
    title style={font=\footnotesize},
    legend style={font=\scriptsize, draw=none, fill=none},
]
\nextgroupplot[
    title={MAE Trend},
    ylabel={Milliseconds},
    legend to name=promptmasklegend,
    legend columns=2,
]
\addplot[thick, mark=*, color=red!75!black] coordinates {
    (21000,27.01) (22000,26.14) (23000,27.03) (24000,23.58) (25000,30.17) (26000,27.51)
};
\addlegendentry{Content MAE}
\addplot[thick, mark=square*, color=blue!70!black] coordinates {
    (21000,18.06) (22000,17.66) (23000,20.51) (24000,17.00) (25000,18.21) (26000,20.56)
};
\addlegendentry{Pause MAE}
\addplot[dashed, thick, color=red!75!black] coordinates {(21000,11.85) (26000,11.85)};
\addlegendentry{Full cond. Content MAE}
\addplot[dashed, thick, color=blue!70!black] coordinates {(21000,9.00) (26000,9.00)};
\addlegendentry{Full cond. Pause MAE}

\nextgroupplot[
    title={Correlation Trend},
    ylabel={Pearson $r$},
]
\addplot[thick, mark=*, color=red!75!black] coordinates {
    (21000,0.694) (22000,0.733) (23000,0.715) (24000,0.773) (25000,0.634) (26000,0.723)
};
\addplot[thick, mark=square*, color=blue!70!black] coordinates {
    (21000,0.553) (22000,0.591) (23000,0.475) (24000,0.543) (25000,0.459) (26000,0.445)
};
\addplot[dashed, thick, color=red!75!black] coordinates {(21000,0.916) (26000,0.916)};
\addplot[dashed, thick, color=blue!70!black] coordinates {(21000,0.769) (26000,0.769)};

\nextgroupplot[
    title={Pause F1 Trend},
    xlabel={SFT Step},
    ylabel={F1},
]
\addplot[thick, mark=*, color=red!75!black] coordinates {
    (21000,0.326) (22000,0.336) (23000,0.289) (24000,0.356) (25000,0.277) (26000,0.293)
};
\addplot[thick, mark=square*, color=blue!70!black] coordinates {
    (21000,0.325) (22000,0.345) (23000,0.267) (24000,0.330) (25000,0.241) (26000,0.275)
};
\addplot[dashed, thick, color=red!75!black] coordinates {(21000,0.413) (26000,0.413)};
\addplot[dashed, thick, color=blue!70!black] coordinates {(21000,0.359) (26000,0.359)};

\nextgroupplot[
    hide axis,
]
\node at (rel axis cs:0.5,0.55) {\ref{promptmasklegend}};
\end{groupplot}
\end{tikzpicture}
\caption{Checkpoint trend on the B@150 test set for prompt-duration-masked MAGIC-TTS checkpoints under prompt-duration-free inference. The best point appears around SFT step 24k, but the gain over the corresponding non-masked setting remains mild and still stays well below the full prompt+target conditioning format.}
\label{fig:appendix_promptmask_sweep}
\end{figure*}

{\bf Promptmask Checkpoint Trend.}
To make this trend more explicit, Figure~\ref{fig:appendix_promptmask_sweep} reports the checkpoint trend on the B@150 test set for prompt-duration-masked MAGIC-TTS checkpoints under prompt-duration-free inference as SFT continues. Relative to the corresponding checkpoint without prompt-time-mask training (`T-only' in Table~\ref{tab:appendix_inference_format}), the best point at SFT step 24k yields only a mild improvement, and it still remains substantially worse than the configuration that keeps prompt-side timing conditions (`Full cond.').

\begin{figure*}[!t]
\centering
\small
\begin{tikzpicture}
\begin{groupplot}[
    group style={group size=2 by 1, horizontal sep=1.8cm},
    width=0.45\textwidth,
    height=0.28\textwidth,
    grid=major,
    xmin=800, xmax=36000,
    xtick={2000,10000,20000,30000,36000},
    xticklabels={2k,10k,20k,30k,36k},
    tick label style={font=\footnotesize},
    label style={font=\footnotesize},
    title style={font=\footnotesize},
]
\nextgroupplot[
    title={Content-Gate Magnitude},
    xlabel={SFT Step},
    ylabel={$|\alpha_{\mathrm{content}}|$ (smooth)},
]
\addplot[thick, mark=*, color=red!75!black] coordinates {
    (800,0.0216) (2000,0.0289) (4000,0.0411) (6000,0.0533) (8000,0.0617) (10000,0.0670)
    (14000,0.0734) (18000,0.0775) (20000,0.0789) (24000,0.0819) (30000,0.0851) (36000,0.0879)
};

\nextgroupplot[
    title={Controllability Metrics},
    xlabel={SFT Step},
    ylabel={Content MAE (ms)},
    ylabel style={color=blue!70!black},
    yticklabel style={color=blue!70!black},
    axis y line*=left,
    axis x line*=bottom,
]
\addplot[thick, mark=*, color=blue!70!black] coordinates {
    (800,15.93) (2000,16.60) (4000,17.13) (6000,14.58) (8000,13.78) (10000,11.99)
    (14000,12.08) (18000,11.54) (20000,11.86) (24000,10.58) (30000,10.22) (36000,10.56)
};
\end{groupplot}
\begin{axis}[
    width=0.45\textwidth,
    height=0.28\textwidth,
    at={($(group c2r1.north west)$)},
    anchor=north west,
    xmin=800, xmax=36000,
    xtick=\empty,
    ylabel={Content Corr.},
    ylabel style={color=green!55!black},
    yticklabel style={color=green!55!black},
    axis y line*=right,
    axis x line=none,
    grid=none,
]
\addplot[thick, mark=square*, color=green!55!black] coordinates {
    (800,0.787) (2000,0.823) (4000,0.831) (6000,0.843) (8000,0.874) (10000,0.903)
    (14000,0.903) (18000,0.911) (20000,0.916) (24000,0.924) (30000,0.922) (36000,0.918)
};
\end{axis}
\end{tikzpicture}
\caption{Training dynamics of the content-conditioning branch during SFT under inference with full prompt text and both prompt-side and target-side duration conditions. The smoothed absolute content-gate magnitude grows almost linearly through training, while test-set controllability improves early and then largely converges.}
\label{fig:content_dynamics}
\end{figure*}
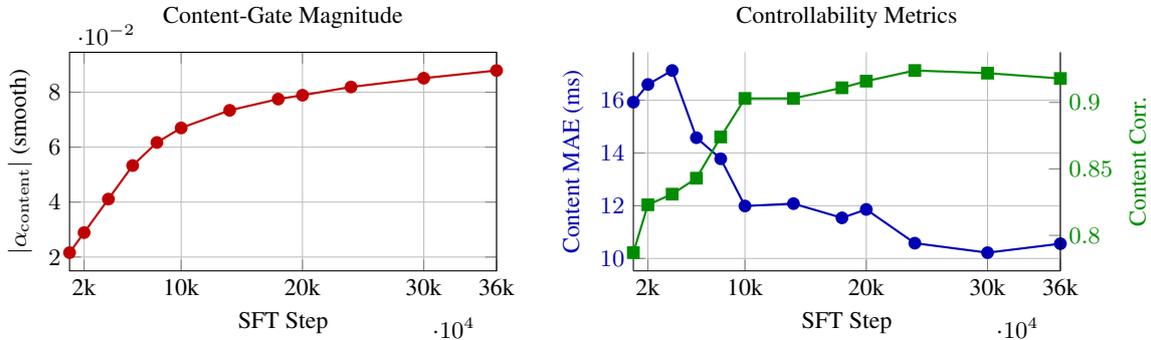

{\bf Content-Gate Dynamics.}
Figure~\ref{fig:content_dynamics} aligns each evaluated timing-control checkpoint with the smoothed content-gate value recorded at the same SFT step. We report smoothed values with smoothing factor 0.6. Since the learned content gate becomes increasingly negative during SFT, we report its absolute value $|\alpha_{\mathrm{content}}|$ as the more relevant proxy for conditioning strength.

The gate trend is close to linear: $|\alpha_{\mathrm{content}}|$ increases from 0.0216 at SFT step 800 to 0.0879 at SFT step 36k. In contrast, timing-control performance improves rapidly in the earlier stage and then largely converges in the late stage, rather than continuing to strengthen in proportion to the gate growth. This analysis shows that the internal strength of the content-conditioning pathway keeps increasing even after downstream test-set controllability has entered a saturation regime. Therefore, the gate increase cannot be read as a complete proxy for controllability on the generalization domain; part of the later growth may instead reflect increasing specialization to the training distribution.

\end{document}